\documentclass[prl, twocolumn, showpacs]{revtex4-1}
\usepackage{graphicx}
\usepackage{amssymb}

\begin{document}
   
\title{Crack front \'echelon instability in mixed mode fracture of a strongly
nonlinear elastic solid }

\date{\today}

\author{O. Ronsin, C. Caroli, T. Baumberger}

\affiliation{ INSP, Universit\'e Pierre et Marie Curie-Paris 6,
CNRS, UMR 7588, 4 place Jussieu, 75005 Paris, France}

\begin{abstract}
Mixed mode (I+III) loading induces segmented crack front \'echelon structures
connected by steps. We study this instability in a highly deformable,
strain-hardening material. We find that \'echelons develop beyond a finite,
size-independent mode mixity threshold, markedly growing  with energy release
rate. They appear via nucleation of localized helical front distortions, and
their emergence is the continuation of the mode I cross-hatching instability of
gels and rubbers, shifted by the biasing effect of shear. This result, at odds
with the direct bifurcation predicted by linear elastic fracture mechanics, can
be assigned to the controlling role of elastic nonlinearity.  

 \end{abstract}

\maketitle

The question of the shape selection of a crack front propagating into a solid --
hence of the resulting crack surface morphology -- while central to fracture
mechanics, remains up to now largely elusive. A striking example is that of
fracture under (I+III) mixed mode loading. While mode I (pure tension) crack
surfaces are basically planar, they generally develop, under superimposed
antiplane (mode III) shear loading\cite{Lawn}, an "\'echelon"
structure\cite{Hull}. This can be roughly described as resulting from a crack
front shape composed of a set of  rotated segments connected by steep steps.
Such \'echelon patterns are quite ubiquitous among materials
: observations range from hard (glasses
\cite{Sommer}\cite{Ravi}\cite{Lazarusexp}, rocks\cite{Cooke}, metals,...) to
soft (gypsum, cheese \cite{Goldstein}) solids.

Up to now, the few theoretical attempts \cite{Ravi}\cite{Cooke}\cite{LL2001}
based on standard tools of linear elastic fracture mechanics (LEFM) and an
additional heuristic ansatz have not been able to predict satisfactorily such
structural features as the rotation of the front segments or their spatial
extension. Besides, they left untouched the issue of  the existence (reported by
Sommer\cite{Sommer}, but never confirmed since) of a finite threshold amount of
mode mixity  for the emergence of \'echelon cracks. In this respect, the recent
work of Pons and Karma  (PK)\cite{PK}, based on a phase field model of brittle
fracture,  constitutes an important opening. They are able to describe their
results on crack propagation as a linear instability of the straight front
against helical deformations, which evolve via coarsening toward the facetted
shape. This has led Leblond, Karma and Lazarus (LKL)\cite{LKL} to perform a
linear stability analysis in the frame of LEFM. On this basis, they predict the
existence of a finite threshold $\left( K_{III}/K_{I}\right)_c$ (where
$K_{I,III}$ are the stress intensity factors imposed by the external loading)
below which the planar crack remains stable.  Moreover, the value of this
threshold is fixed by that of the Poisson ratio only, and does not depend on the
energy release rate. However, the absence of any length scale in the LEFM
framework results in a pathological feature of the corresponding bifurcation:
for $\left( K_{III}/K_{I}\right) >  \left( K_{III}/K_{I}\right)_c$, all
wavelengths $ \lambda$ become simultaneously unstable, and the growth rate of
the front distortion diverges as  $\lambda \rightarrow 0$. As pointed by LKL,
tackling the associated regularization calls for identifying a small length
cutoff, the physical origin and degree of material dependence of which remain
open issues.

With this question in mind, we study here (I+III) mixed mode fracture in a
gelatin gel, the choice of this material being dictated by its mechanical
specificities. Indeed, in this physical hydrogel, fracture proceeds via
stress-induced unzipping of (triple helix) crosslinks of the gelatin network,
and subsequent dissipative pull-out of the unzipped polymer chains \cite{nous}.
This demands that, in the crack tip vicinity, stresses build up to a level $\sim
100 E$ , with $E$ the small strain Young modulus. As discussed by Hui
\cite{Hui}, such a level of stress concentration, while prohibited in linear
elastic materials by elastic blunting, can be reached in soft polymer gels
thanks to the huge strain hardening signaling the crossover between the coiled
(entropic) and taut chains (enthalpic) elastic regimes. The extension of the
near-tip region where non linearities become relevant naturally provides a small
scale cutoff $\ell_{NL}$ below which the universal inverse square root LEFM
stress divergence no longer holds, as directly demonstrated by Livne {\it et al}
\cite{Livne}.   Estimating this length as the distance to the tip where the LEFM
stress reaches a value $\sim E$ leads, in agreement with ref.\cite{Geubelle}, to
$\ell_{NL} \sim \mathcal G /E$, with $\mathcal G$ the fracture energy. In
hydrogels $\ell_{NL}$ typically lies in the $100 \mu$m--$1$mm range
\cite{magic}\cite{Bouch}, much larger than both the process zone and network
mesh sizes. It can thus be expected to play a decisive role in crack path
selection. Indeed, it has been shown to govern the oscillatory instability in
rapid fracture of brittle gel films \cite{Bouch} as well as the crack branching
response of gelatin gels to a solvent-induced  environmental shock
\cite{branching}.

 Extensive exploration of crack surface morphologies in the quasi-static regime
reveals the existence in our system of a finite mode mixity threshold which, at
variance with the LKL prediction, strongly increases with $\mathcal G$.
Moreover, we bring direct evidence that the emergence of the \'echelon
structures does not follow the linear-instability-plus-coarsening PK picture.
Rather, they develop via a mechanism akin to the emergence of the cross-hatching
(CH) instability observed in mode I fracture of gels \cite{magic}\cite{Seki} and
rubbers \cite{Gent} -- namely, the localized nucleation of steps triggered by
structural fluctuations whose minimum amplitude decreases as $K_{III}/K_{I}$
grows.  From this we conclude that, in the presence of large elastic crack tip
blunting, the mixed mode fracture response is fully controlled by
non-linearities. We then argue that, more generally, the relevance of the LKL
linear stability analysis should be correlated with the amplitude of mode I
crack surface roughness.

\begin{figure}[htbp]
\begin{center}
\includegraphics[width=4.5cm]{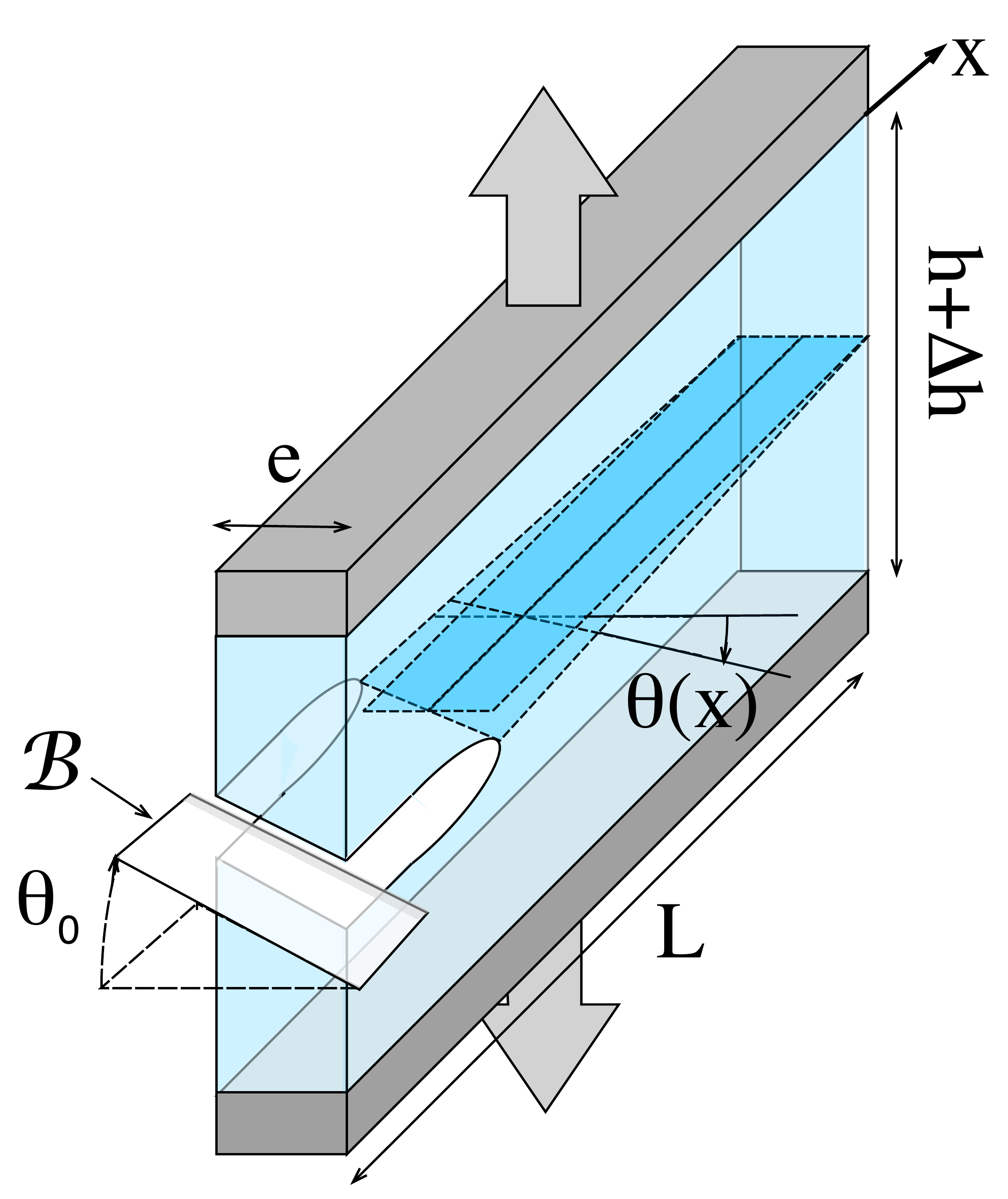}
\caption{Sketch of the experimental setup: the gel slab (unstrained dimensions
$L\times h \times e$), once  stretched by $\Delta h$, is notched at angle
$\theta_0$ by the blade $\mathcal B$.
}
\label{fig:Setup}
\end{center}
\end{figure}

 Our experiments are performed, according to the protocol described in ref
\cite{EPJE}, on gel slabs ($E = 12$kPa) made of $5$ wt$\%$ gelatin in a
water-glycerol mixture. Sample sizes and compositions are listed in Table
\ref{tab:samples}.
Mixed mode loading is introduced by notching the sample, previously stretched to
the desired level, at an angle $\theta_{0}$ (see Fig.\ref{fig:Setup}) from the
plane of propagation of pure mode I cracks.
Profilometric anaysis of crack surfaces is performed as described in the
Supplemental Material  \cite{Supp}.  

We choose for our control parameters
(i) the mode mixity indicator $ m_{0} = \tan \theta_{0}$ (which would measure
$K_{III}/K_{I}$ in the $e \rightarrow \infty$ limit), with $\theta_{0}$ ranging
up to 50 deg. (ii) the initial energy release rate $\mathcal G = \mathcal W
(Le/cos \theta_{0})^{-1}$, with $\mathcal W$ the loading work. Fracture of
gelatin being highly dissipative, in the explored loading range ($\mathcal G <
20$ J.m$^{-2}$), mode I crack propagation is steady and quasistatic (velocities
$V \lesssim$ a few mm.s$^{-1}$)\cite {nous}.

\begin{table}[htdp]
\caption{Gel sample characteristics. Slab dimensions (in cm) as defined on
Fig.\ref{fig:Setup}.  $\eta$ (in mPa.s) is the viscosity of the water-glycerol
solvent.  }
\begin{center}

\begin{tabular}{|c|c|c|c|c|}
\hline
{\it slab symbols}&$L$&$h$&$e$&$\eta$\\

\hline
diamonds&30&\,3\,&\,1\,&\,11\,\\
\hline
squares&\,30\,&3&1&3.6\\
\hline
circles&30&3&2&11\\
\hline
down triangles&30&10&1&11\\
\hline
up triangles&20&2&0.5&11\\
\hline
\end{tabular}
\end{center}
\label{tab:samples}
\end{table}

 \begin{figure}[htbp]
\begin{center}
\includegraphics[width=8.5 cm]{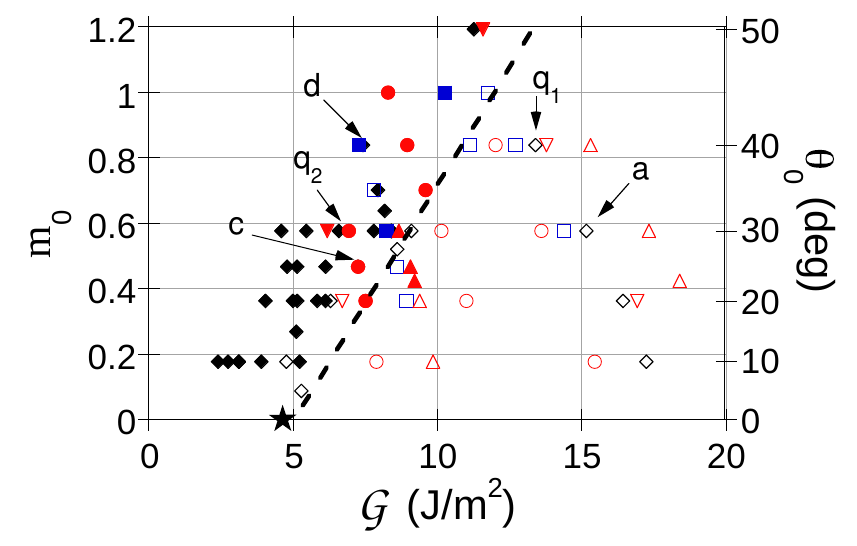}
\caption{Crack surface morphology, in the (energy release rate $\mathcal G$,
mode mixity indicator $m_0$)  control parameter plane for samples with various
sizes and solvent compositions. Symbol shapes are defined in Table
\ref{tab:samples}.
Empty symbols: smooth cracks; Full symbols: \'echelon cracks. The dashed line
defines the \'echelon instability threshold $m_c(\mathcal G)$. The star symbol
corresponds to the onset of the mode I cross-hatching instability (see text).
Labels a, c, d refer to Figure \ref{fig:patterns}. For q$_1$, q$_2$ see text.
}
\label{fig:diagram}
\end{center}
\end{figure}

Our results on crack surface morphologies are summarized on Fig.\ref
{fig:diagram}, which gathers data from the ($L,e, h, \eta$) sets listed in Table
\ref{tab:samples}. In agreement with previous works, we observe that mode III
induces echelon cracks. However, we find that they only appear beyond a finite
mode mixity threshold $m_{c}$ (dashed line on Fig.\ref{fig:diagram}), thus
confirming Sommer's suggestion~\cite{Sommer}. Strikingly, the domain of
existence of \'echelon fronts in the ($m_{0}, \mathcal G$) parameter space turns
out to be insensitive to both the slab geometry and the solvent viscosity.
However, $m_{c}$ is not a constant but grows markedly with $\mathcal G$. As
$m_{0}$ is increased, the fracture behavior evolves as follows.

\begin{figure}[htbp]
\begin{center}

\includegraphics[width=8.5cm]{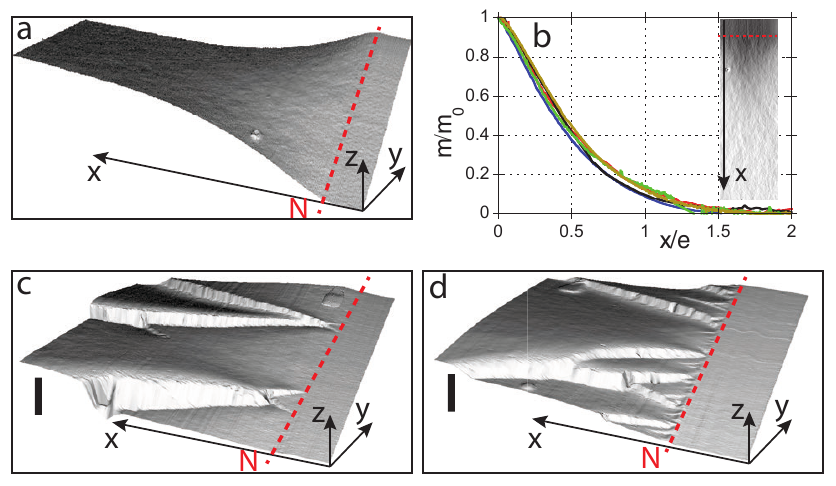}

\caption{Crack surface morphologies for parameter values as labeled on Fig
\ref{fig:diagram}. (a): Surface profile of a smooth crack (area $6 \times 24$
mm$^{2}$). (b) Scaled mode mixity parameter ($m/m_0$) {\it vs}  scaled
propagation distance ($x/e$) from notch line N, for: $\{e = 1$ cm, $m_0 = 0.27,
0.36, 0.58, 0.84\}$ and $\{e = 2$ cm, $m_0 = 0.36\}$. {\it Inset:
}Microroughness of crack (a) as revealed by grazing angle illumination. (c,d):
Surface profiles of cracks
close beyond threshold (c), and deeper into the \'echelon regime (d) (areas
$8Ê\times 10$ mm$^{2}$, vertical bar $500 \mu$m). }
\label{fig:patterns}
\end{center}
\end{figure}

{\it Smooth regime:} At small notch angles ($m_{0} < m_{c}$), the crack front
remains quasi-linear \cite{Supp}.
However, as it propagates, its tilt angle $\theta(x)$ decreases steadily until
it returns to the mode I configuration. Indeed, we meet here with the difficulty
inherent to mixed mode fracture experiments - namely it is in practice
impossible to impose a steady mode mixity amount. Here, as in refs
\cite{Ravi}\cite{Lazarusexp}, we only impose its initial value. A crack surface
typical of this smooth regime is shown on Fig.\ref{fig:patterns}.a. Profile
analysis for samples with various $\theta_{0}$ and $e$ values (see
Fig.\ref{fig:patterns}.b) shows  that the tilt angle relaxation curves obey the
scaling law: $m(x) = m_{0} f(x/e)$, where $f$ exhibits an exponential-like decay
over a range $x/e \simeq 1$: in agreement with Saint-Venant's principle, the
memory of initial conditions is constrained by the sample
geometry\cite{footnote2}. Finally, closer examination reveals (insert of
Fig\ref{fig:patterns}.b), on top of the global smooth profile, a roughness of
micrometric r.m.s. amplitude with the same characteristics as that of mode I
crack surfaces.

{\it Echelon regime:} For $m_{0} \gtrsim m_c(\mathcal G)$
(Fig.\ref{fig:patterns}.c), the fracture surfaces exhibit highly localized
steps, oriented so as to reduce the tilt  of the facets which join them. They
nucleate at the very tip of the notch and grow, as the crack propagates, up to
heights $\sim 100 \mu$m. Moreover, we find that they systematically drift along
the front, in a seemingly random direction.  As $m_0$ gets larger (Fig
\ref{fig:patterns}.d), the step density increases and we observe step
``collisions" which result in their merging (Fig \ref{fig:patterns}.a,b). Thanks
to the drift, the steps successively exit from the slab, thereby relaxing the
average tilt angle and finally leaving a smoothed crack surface.

Let us now put this set of results in regard to the instability scenario
emerging from refs.\cite{PK}\cite{LKL}.  While we  do agree with LKL on the
existence of a finite mode mixity threshold, we notice two main discrepancies:

(i) At variance with their prediction of a $\mathcal G$-independent $m_c$ value
($\simeq 0.23$ for our quasi-incompressible gel) we measure a tenfold increase
when $\mathcal G$ grows by a factor $\sim$ 2. Note that, since we find that
$m_c(\mathcal G)$ is independent of sample thickness, this difference cannot
originate from a mere finite size effect. 

(ii) PK suggest that \'echelon patterns emerge via a linear instability as a
quasi sinusoidal front modulation which then coarsens as it amplifies, according
to a direct bifurcation picture. We observe a markedly different behavior,
namely nucleation and growth of highly localized front distortions.

\begin{figure}[htbp]
\begin{center}
\includegraphics[width=8.5cm]{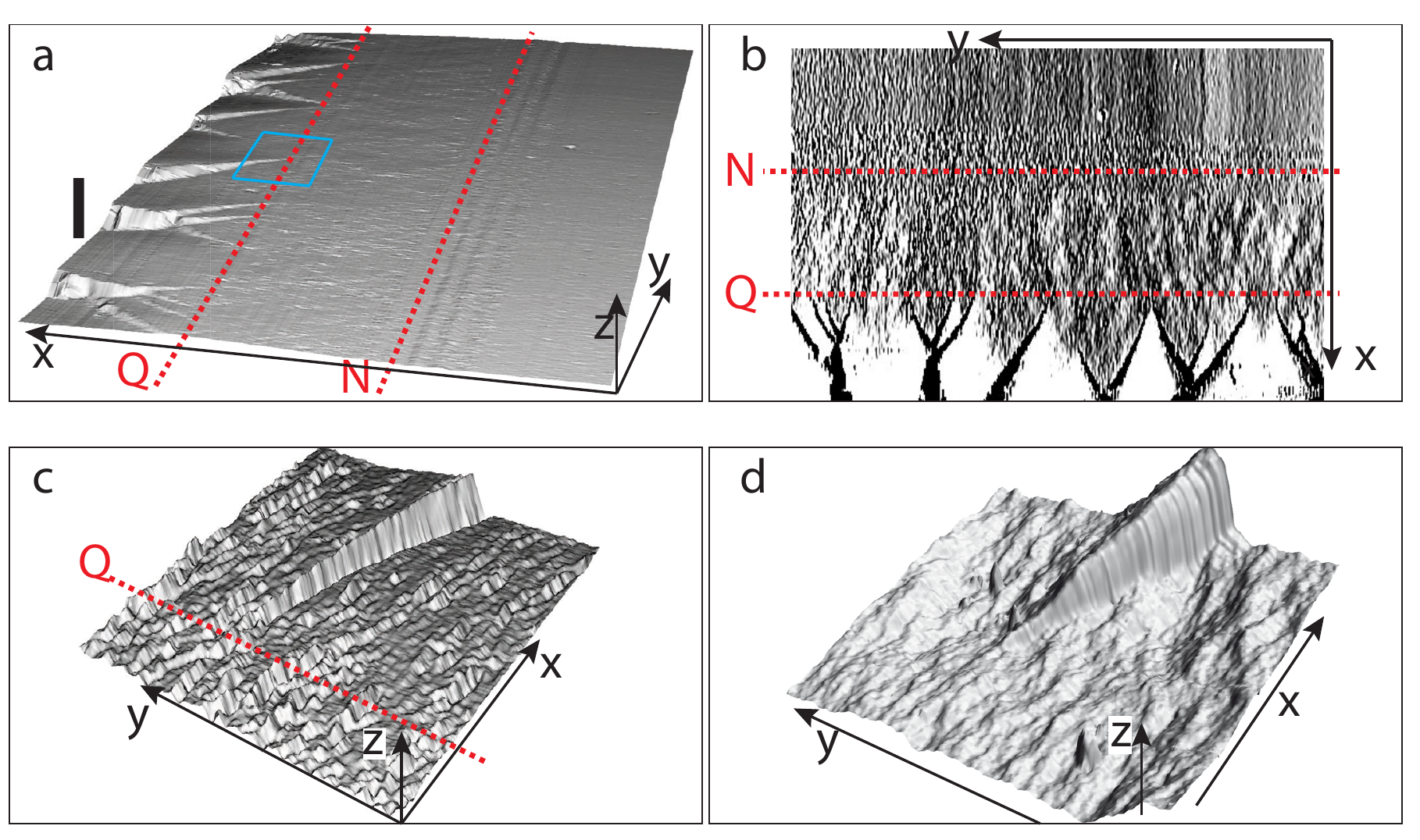}
\caption{(a): Surface profile corresponding to the quench protocol (see text): N
indicates the notching line, Q corresponds to the position reached by the smooth
crack front at the time of the mechanical quench (area $5 \times 3.3$ mm$^2$,
vertical bar 300 $\mu$m). (b) Same data plotted under simulated grazing
illumination showing grooves left by the notching blade (above N), smooth crack
microroughness (region NQ) and \'echelon emergence (below Q). White regions are
facing the light source. (c) Blow up of the square area marked on (a), showing
nucleation and growth of an \'echelon step (area $500 \times 500 \mu$m, peak-to
peak 20 $\mu$m). (d) (Adapted from Baumberger {\it et al.} \cite{magic}) Nucleation
and growth of  a cross-hatching defect emerging on the surface of a mode I crack
(area $400 \times 400 \mu$m, peak-to peak 15 $\mu$m).
}
\label{fig:quench}
\end{center}
\end{figure}

 These discrepancies raise an obvious issue: could nucleation, in  our
experiments, be heterogeneous, i.e. triggered by defects imprinted by the
notching blade? In order to address this question, we have performed the
following ``mechanical quench" experiment: we initiate a crack in the smooth
regime (point $q_1$ on Fig.\ref{fig:diagram}),
let it propagate over a distance of 1.2 mm, then suddenly reduce the remote
tensile loading, thus bringing the crack into the \'echelon regime (point
$q_2$). Fig \ref{fig:quench}.a shows the corresponding surface profile.
Unambiguously, the steps nucleate quasi-simultaneously and without measurable
delay on the quench line $Q$. Moreover, as illustrated on the blow-up Fig
\ref{fig:quench}.c, they emerge from ridges of the micro-roughness already
present in the pre-quench, nominally smooth, $NQ$ region. On the other hand, we
find (see Fig.\ref{fig:quench}.b) that this micro-roughness exhibits no
discernible correlation with the blade-generated grooves. We therefore conclude
that the \'echelon instability develops via homogeneous nucleation of localized
structures.

This phenomenology is highly reminiscent of that of the Cross-Hatching (CH)
instability previously observed in mode I fracture of highly deformable
materials and analyzed, on the case of gelatin,  in ref.\cite{magic}. This
latter instability is characterized by the emergence, below a critical energy
release rate $\mathcal G_{CH}$, at random locations across the slab width, of
narrow steps drifting in either direction. We have interpreted them as the
response of the crack front to transient pinning by a toughness fluctuation:
this gives rise to a local (I+II+III) mixed mode configuration, hence to a 
helicoidal self-amplifying front deformation evolving into two connected
half-cracks. The resulting, symmetry-breaking, local morphology is thus akin to
the (I+III) \'echelon one, while the global symmetry imposed by the absence of
remote shear loading reflects into the random sign of the facet tilt angles.

As immediately appears when comparing Figs.\ref{fig:quench}.c,d, \'echelon and
CH steps develop in a strikingly similar way. This suggests that, in our soft
elastic system, the \'echelon instability might be the continuation of the CH
one, that would be simply shifted under the biasing effect of shear loading.
If such is the case, the \'echelon threshold line $m_c(\mathcal G)$ must
extrapolate to $G_{CH}$ in the $m_0 = 0$ (pure mode I) limit. For the gels
studied here, we measure a  $G_{CH}$ value of  $4.6 \pm 0.3$ J.m$^{-2}$, in
agreement with \cite{magic}. As seen on Fig.\ref{fig:diagram}, we find that this
value is fully compatible with the extrapolation of the \'echelon threshold
line.  Let us moreover note that $m_c(\mathcal G)$ turns out to be  insensitive
to a threefold variation of the solvent viscosity $\eta$, in agreement with the
$\eta$-independence of $G_{CH}$ \cite{magic}.

In summary, we are able to assert that, as far as the \'echelon instability is
concerned, the effect of a finite mode mixity amount is merely to facilitate the
local development of  helical defects generated on front pinning tough sites
provided by the intrinsic disorder of the random polymer gel network.

This conclusion may at first sight appear quite puzzling. Indeed, it points to
the irrelevance to our system of the linear instability-plus-coarsening
scenario. If such is the case, why is the (formally indisputable) LKL linear
analysis not valid here or, in other words, what is the criterion for its
validity?

With regard to this question, it is important to note that, as shown in
\cite{magic}, the in and out of plane front deviation amplitudes for a fully
developed CH defect turn out to be on the order of the cutoff length $\ell_{NL}$
associated with elastic non-linearities. In other words, the nucleation and
initial development of such defects take place in the elastically blunted, fully
strain-hardened, tip vicinity. Hence the fact that LEFM cannot account for these
processes.

Nevertheless, as discussed above, the LKL approach should be legitimate on
length scales larger than $\ell_{NL}$, and the spatially extended
destabilization which it predicts might compete with the biased CH process. Let
us stress, however, that this extended response must be understood as driven by
the noise {\it renormalized up to the coarse-graining scale}. Now, as the LEFM
front problem is itself non-linear, the front deformation as calculated by LKL
is the first term of an amplitude expansion, and truncation to first order is
valid only for noise amplitudes below some cutoff $A_{max}$ fixed by the precise
expression of next order terms. Note that $A_{max}$ should exhibit a universal
dependence on the small strain elastic coefficients. On the other hand,
the coarse-grained noise amplitude scale $\bar A(\ell_{NL})$, which can in
principle be evaluated from the micro-roughness spectrum of mode I cracks, 
certainly depends on nonlinear (elastic, plastic) material properties.  We
believe our strongly nonlinear elastic system to correspond to the regime  $\bar
A(\ell_{NL}) > A_{max}$ in which the  extended instability  is irrelevant.

These remarks clearly suggest the validity of the LKL scenario  to be material
dependent and, hence,
the nature of the \'echelon instability itself  to be a non-universal feature.
Insofar as calculating higher orders terms of the LEFM amplitude expansion
appears quite a formidable task, we hope that it will be possible to test  this
proposition by extending the phase field approach to include nonlinear
elasticity.

\acknowledgements

We are grateful to A. Karma, V. Lazarus, K. Ravi-Chandar for stimulating
discussions and to H. Henry for communication and discussion of unpublished
results.

\newpage

~

\newpage

\onecolumngrid

\centerline{\bf\large Supplemental Material:}
\smallskip
\centerline{\bf\large Crack front \'echelon instability in mixed mode fracture of a strongly nonlinear elastic}
\smallskip
\centerline{\bf\large solid}

\bigskip

\centerline{O. Ronsin, C. Caroli, T. Baumberger}

\bigskip

\bigskip

\paragraph*{\bf Profilometric analysis of crack surfaces:}

In order to circumvent sample evolution induced by solvent evaporation, we make replicas of the fresh fracture surfaces by casting and UV-curing thin layer of glue. The replicas are fixed on a two-axes x-y translation stage and scanned with the help of a confocal chromatic optical sensor (Micro Epsilon NCDT IFS 2401-1). The sensor itself is fixed to a z translation stage so as to extend its nominal 1 mm measuring range up to 2 cm. This home-built  profilometer enables us to scan zones of typical areas $1\times 5$ cm$^2$ with a  vertical resolution of  100 nm and a lateral one of $10\times10 \,\mu$m$^2$. Note however that, due to the need for glue trimming along the replica edges, the scanning zone only extends up to a distance of 1mm from each edge.    

While the tilt angle $\theta(x)$ is defined as the inclination of the front line in the   stretched sample, with respect to the sample (horizontal) mid-plane, profile measurements are performed on the unloaded surfaces. In order to identify the  $\theta =0$ reference plane, all experimental runs are continued up to propagation distances such that the crack has returned to the mode I configuration. From the value $\alpha(x)$ of the   measured tilt angle we infer $\theta(x)$ according to:
$$\theta(x) = \tan^{-1}\left [\left(1+\frac{\Delta h}{h}\right )^2\tan\alpha (x)\right ]$$
which accounts for the affine deformation of the incompressible slab due to the grip displacement $\Delta h$. We have checked that we thus recover the value $\theta_0$ of the notching angle up to $\pm 5\%$.\\

\paragraph*{\bf Profiles in the smooth regime: }
Figure S$_1$ shows sections $h(y)$ of  a ``smooth'' fracture surface by planes perpendicular to the propagation direction $x$. As analyzed in reference [Bisen Lin, M.E. Mear and K. Ravi-Chandar, Int. J. Frac., {\bf 165}, 175 (2010)],  finite slab thickness induces a mode II loading component which decreases gradually from the edges and vanishes on the mid-plane $y=0$.  Although we cannot access the immediate vicinity of the edges (see above), one indeed clearly distinguishes on Figure S$_1$  the bending trend expected to result from this effect.   We choose to define the tilt angle from the slope of the profile at $y=0$.  
We then compute with the help of the above equation  the local mode mixity indicator $m(x) = \tan \theta(x)$, whose evolution is  shown on Fig. 3b for cracks in the smooth regime.

\begin{center}
\includegraphics[scale=0.5]{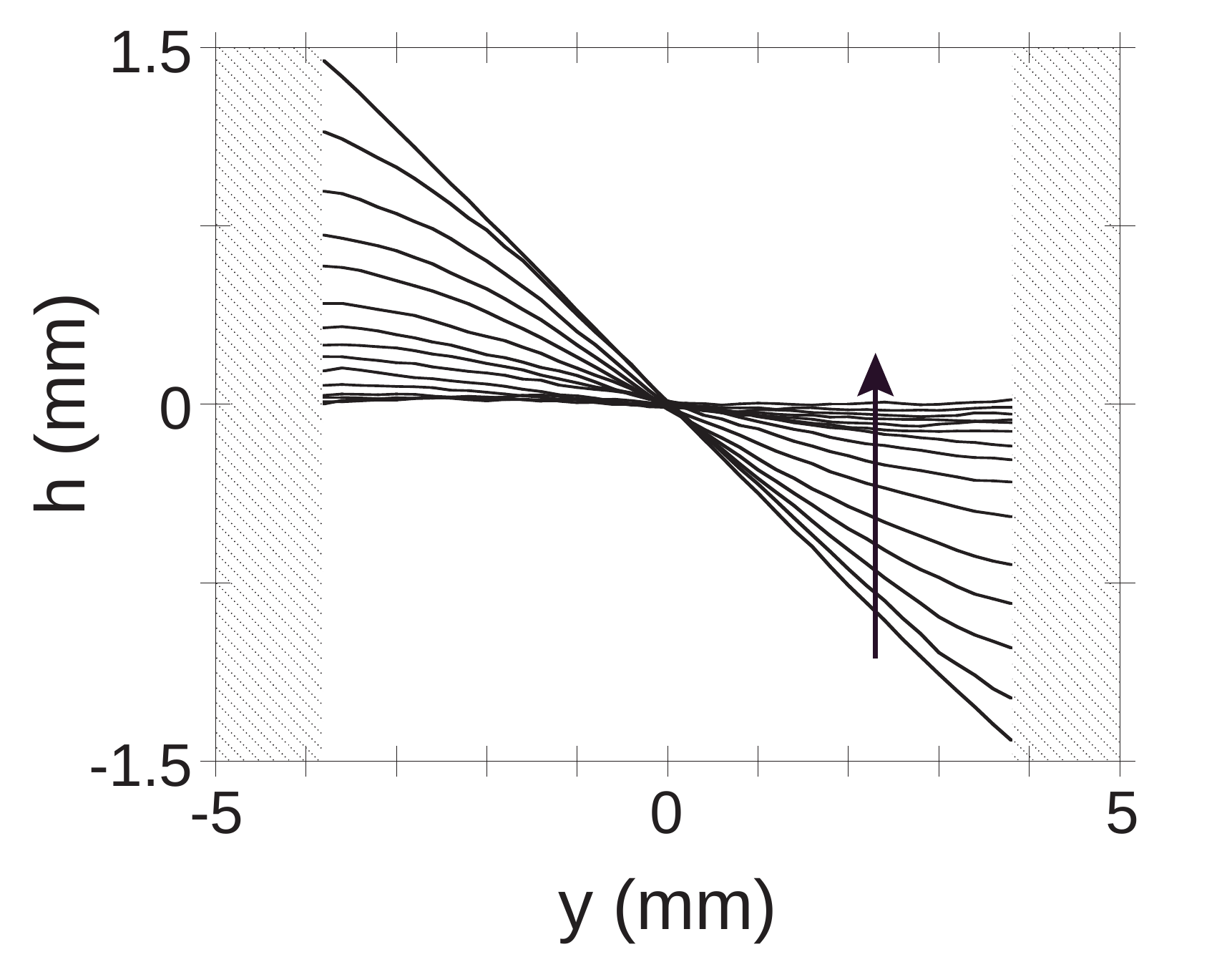}
 \end{center}
 {\small Fig. S$_1$: Sections of the smooth fracture surface shown on Fig. 3a, by planes perpendicular to the propagation direction $x$ and  evenly spaced ($\Delta x= 2$ mm) from the initial straight notch, growing in the direction of the arrow. The  hatched stripes indicate the edge regions of the $10$ mm wide sample inaccessible to scanning.  }

\end{document}